\documentclass[numreferences]{kluwer}
\usepackage{amsmath,amssymb,amsthm}
\numberwithin{equation}{section}

\newcommand{\zb}{\boldsymbol{b}}

\newcommand{\zh}{\boldsymbol{h}}

\newcommand{\zl}{\boldsymbol{l}}

\newcommand{\zn}{\boldsymbol{n}}

\newcommand{\zx}{\boldsymbol{x}}
\newcommand{\zy}{\boldsymbol{y}}

\newcommand{\zjm}{\boldsymbol{\jmath}}
\newcommand{\zell}{\boldsymbol{\ell}}

\newcommand{\zC}{\boldsymbol{C}}
\newcommand{\zD}{\boldsymbol{D}}

\newcommand{\zQ}{\boldsymbol{Q}}
\newcommand{\zS}{\boldsymbol{S}}
\newcommand{\zT}{\boldsymbol{S}}

\newcommand{\zZ}{\boldsymbol{Z}}

\newcommand{\zcB}{\mathcal{B}}

\newcommand{\zcP}{\mathcal{P}}

\newfont{\mmit} {cmmi10 scaled 1200}%
\newfont{\smit} {cmmi7 scaled 1200}%
\newfont{\ssmit}{cmmi5 scaled 1200}%

\newfont{\tenbfit}{cmmib10}%
\newfont{\svnbfit}{cmmib10 scaled 800}%
\newcommand{\zSigma}{\mbox{\tenbfit\char'06\/}}%
\newcommand{\szSigma}{\mbox{\svnbfit\char'06\/}}%

\newfont{\tenbfsl}{cmbxti10}
\newcommand{\idem}{\mbox{\tenbfsl 1\/}}%

\newcommand{\zxi}{\boldsymbol{\xi}}

\newcommand{\zcdot}{\cdot}   
\newcommand{\zdot}[1]{\skew1\dot{\mathnormal{#1}}}
\newcommand{\zzdot}[1]{\skew3\dot{\mathnormal{#1}}}

\newcommand{\phat}[1]{{\skew3\hat{#1}}} 
\newcommand{\xg}{\textsl{Grad\:}}   
\newcommand{\xd}{\textsl{Div\:}}
\newcommand{\Det}{\textsl{Det}\:}

\newcommand{\da}{\:\textsl{da}}
\newcommand{\dv}{\:\textsl{dv}}

\newcommand{\kB}{{k_{\rm B}}}           
\newcommand{\kBT}{{k_{\rm B}\theta}}    

\newcommand{\half}{\hbox{$\frac{1}{2}$}} 

\newcommand{\trans}{\scriptscriptstyle\top\mskip-4mu} 

\begin{document}
\begin{opening}

\title{Theory for atomic diffusion on fixed and deformable crystal lattices}
\author{Eliot Fried}
\institute{Department of Theoretical and Applied Mechanics
\\University of Illinois at Urbana-Champaign
\\Urbana, IL 61801-2935, USA}
\author{Shaun Sellers}
\institute{School of Mathematics
\\University of East Anglia
\\Norwich NR4 7TJ, England}
\date{}
\runningtitle{Atomic diffusion on fixed and
deformable crystal lattices}
\runningauthor{E.\ Fried \& S.\ Sellers}

\dedication
{Dedicated to Roger Fosdick, whose fundamental and broad-reaching work in
continuum thermomechanics has provided us with abundant inspiration.}

\begin{abstract}
We develop a theoretical framework for the diffusion of a single
unconstrained species of atoms on a crystal lattice that provides a
generalization of the classical theories of atomic diffusion and
diffusion-induced phase separation to account for constitutive
nonlinearities, external forces, and the deformation of the lattice. In this
framework, we regard atomic diffusion as a microscopic process described by
two independent kinematic variables: ({\em i\/}) the atomic flux, which
reckons the local motion of atoms relative to the motion of the underlying
lattice, and ({\em ii\/}) the time-rate of the atomic density, which
encompasses nonlocal interactions between migrating atoms and characterizes
the kinematics of phase separation. We introduce generalized forces
power-conjugate to each of these rates and require that these forces satisfy
ancillary microbalances distinct from the conventional balance involving the
forces that expend power over the rate at which the lattice deforms. A
mechanical version of the second law, which takes the form of an energy
imbalance accounting for all power expenditures (including those due to the
atomic diffusion and phase separation), is used to derive restrictions on
the constitutive equations. With these restrictions, the microbalance
involving the forces conjugate to the atomic flux provides a generalization
of the usual constitutive relation between the atomic flux and the gradient
of the diffusion potential, a relation that in conjunction with the atomic
balance yields a generalized Cahn-Hilliard equation.
\end{abstract}

\end{opening}

\section{Introduction}
\label{subsec:diffusion}

The diffusion of a single unconstrained atomic species on a fixed crystal
lattice is commonly modeled by the Cahn-Hilliard equation (see, for example,
Cahn \cite{C1,C2,C3} and Cahn \& Hilliard \cite{CH1,CH2,CH3})
\begin{equation}
\zzdot{\nu}=\xd\bigl(D_1(\nu)\, \xg\nu-D_2(\nu)\,\xg(\xd\xg\nu)\bigr),
\label{eq:cahn-hilliard}
\end{equation}
with $\nu$ the \emph{atomic density}, $D_1$ the nonnegative
density-dependent \emph{diffusivity}, and  $D_2$ a small, nonnegative
density-dependent coefficient. The first term on the right side is
due to simple diffusion, whereas the second one is a higher order term that
takes into account spatial inhomogeneities on a fine scale, such as those
that may arise during processes involving phase separation.
This equation can be derived by adjoining to the \emph{atomic balance}
\begin{equation}
\zzdot{\nu}=-\xd\zjm,
\label{eq:mass}
\end{equation}
the constitutive relation
\begin{equation}
\zjm=-\kappa(\nu)\, \xg\mu,
\label{eq:constitutive1}
\end{equation}
determining the \emph{atomic flux} as proportional to the gradient $\xg\mu$ of
the \emph{diffusion potential} $\mu$ through a nonnegative scalar
\emph{mobility} $\kappa$ that, in general, depends on the atomic density. In
\emph{simple atomic diffusion}, the diffusion potential is given by the
derivative of a constitutive response function $\hat{\psi}$ that determines
the \emph{energy density} $\psi$ in terms of the atomic density, viz.,
\begin{equation}
\mu=\hat{\psi}{}^\prime(\nu),
\end{equation}
in which case the diffusive atomic flux takes the form
%
\begin{equation}
\zjm=-D(\nu)\,\xg\nu,
\label{eq:constitutive2}
\end{equation}
with diffusivity $D$ determined by
\begin{equation}
D(\nu)=\kappa(\nu)\hat{\psi}^{\prime\prime}(\nu).
\end{equation}
However, when strong spatial inhomogeneities in the density may be
present, the response function determining the energy density is taken to
depend on both the density $\nu$ and its gradient $\xg\nu$, so that the
diffusion potential is given by the \emph{variational derivative}
\begin{equation}
\mu=\frac{\partial\hat{\psi}}{\partial\nu}(\nu,\xg\nu)
-\xd\bigl(\frac{\partial\hat{\psi}}{\partial(\xg\nu)}(\nu,\xg\nu)\bigl).
\label{eq:mu}
\end{equation}
Often, in this context, it is presumed that $\hat{\psi}$ has the particular
form
\begin{equation}
\hat{\psi}(\nu,\xg\nu)=f(\nu)+\half\alpha\, |\xg\nu|^2,
\end{equation}
with $\alpha$ a positive constant that is usually interpreted as
proportional to a squared characteristic length of interaction and, hence,
taken to be small, so that the atomic flux becomes
\begin{equation}
\zjm=-D_1(\nu)\,\xg\nu+D_2(\nu)\,\xg(\xd\xg\nu),
\end{equation}
with coefficients
\begin{equation}
D_1(\nu)=\kappa(\nu)f^{\prime\prime}(\nu)
\qquad\text{and}\qquad
D_2(\nu)=\alpha\kappa(\nu).
\end{equation}

Recently, Gurtin \cite{G} presented a thermodynamically consistent framework
that provides a broad generalization of (\ref{eq:cahn-hilliard}) that
accounts for constitutive nonlinearities, kinetics, and deformation  of the
underlying lattice. This framework is based on the belief  that to each
independent kinematic process there should correspond a system of
power-conjugate forces subject to a generalized force balance.\footnote{A
foundation for this viewpoint can be devised using the {\em principle of
virtual power}, see, for example, Germain \cite{Germain}, Antman \& Osborn
\cite{Antman}, Maugin \cite{Maugin}, and Fr\'emond \cite{Fremond}. We follow
here an alternative, but essentially equivalent, approach taken by Ericksen
\cite{E61}, Goodman \& Cowin \cite{GC72}, Capriz \& Podio-Guidugli
\cite{CP-G}, Capriz \cite{C89}, Fried \& Gurtin \cite{FG1,FG2},  Fried
\cite{F}, and Gurtin \cite{G}.} Specifically, Gurtin \cite{G} identifies
atomic diffusion and lattice deformation as distinct kinematical processes,
with the density-rate $\zzdot{\nu}$ being a generalized velocity associated
with atomic diffusion and the deformation-rate $\skew3\dot{\zy}$ being the
velocity of the underlying lattice. Associated with $\zzdot{\nu}$ is a
system of generalized forces that satisfy a scalar \emph{microforce
balance}. This balance is a postulate distinct from and ancillary to the
conventional force balance involving forces that act power conjugate to the
velocity of the underlying lattice. With appropriate thermodynamically
consistent constitutive equations, the microforce balance yields an
expression for the diffusion potential that generalizes (\ref{eq:mu}).
Finally, the generalized Cahn-Hilliard equation arises on combining with the
atomic balance the thermodynamically derived expression determining the
atomic flux as proportional to the gradient of the diffusion potential.

Although this approach provides a powerful and appealing means of deriving
(\ref{eq:cahn-hilliard}) and various generalizations thereof, we believe that
it is lacking in the sense that it accounts only for one aspect of the
kinematics associated with atomic diffusion. In particular, we take the view
that the diffusive flux of atoms, as described by $\zjm$, should be
considered as a kinematic process distinct from that described by the
density-rate $\zzdot{\nu}$. Specifically, we conceive of atomic diffusion as
a microscopic process described by
\begin{itemize}
\item[(\emph{i})] the atomic flux $\zjm$, which measures the local motion of
atoms relative to macroscopic motion of the lattice as characterized by the
velocity $\skew3\dot{\zy}$, and by
\item[(\emph{ii})] the rate $\zzdot{\nu}$ of the atomic density, which
encompasses nonlocal interactions between diffusing atoms and characterizes
the kinematics of phase separation.
\end{itemize}
%
Adopting the view promoted by Fried \& Gurtin \cite{FG1,FG2}, Fried \cite{F},
and Gurtin \cite{G}, we arrive at an alternative derivation of the
Cahn-Hilliard equation (\ref{eq:cahn-hilliard}), where, to account for power
expenditures associated with the microscopic motion of atoms, we introduce
two separate systems of forces---one being power-conjugate to the atomic flux
$\zjm$ and the other being power-conjugate to the rate $\zzdot{\nu}$ of the
atomic density. Further, we require that these forces be consistent with
separate microforce balances, one vectorial and involving the forces
power-conjugate to $\zjm$ and the other scalar and involving the forces
power-conjugate to $\dot{\nu}$.\footnote{Previously, Fried \& Sellers
\cite{FS} developed a theory for solute transport in which the diffusive mass
flux is interpreted as a generalized velocity and corresponding generalized
forces subject to an ancillary force balance are introduced.}

We first focus on the development of a simple theory, where, to highlight the
roles of the vectorial and scalar microforces, the lattice is fixed. Next,
we extend the theory to include deformation of the lattice. For simplicity,
we restrict our attention to processes involving only atomic diffusion and
lattice deformation, ignoring thermal and other effects, so that the first
and second laws of thermodynamics are replaced by an energy imbalance that
accounts for all relevant power expenditures. As a further simplification,
we ignore all forms of inertia. We employ a purely referential description
and consistently use direct notation (see, for example, Gurtin
\cite{gurtin81}).

\section{Simple theory for atomic diffusion on fixed lattice}

\subsection{Balance and imbalance laws}
\label{sect:balance}

The theory is based upon the following laws:
\begin{itemize}
\item[$\bullet$] atomic balance;
\item[$\bullet$] flux-conjugate microforce balance;
\item[$\bullet$] density-rate-conjugate microforce balance;
\item[$\bullet$] energy imbalance.
\end{itemize}
In contrast to standard formulations of diffusion in rigid bodies, we
include in this list of balances \emph{two} microforce balances, which will
be seen to provide appropriate generalizations of commonly assumed
constitutive relations. In particular, the flux-conjugate microforce balance
provides the generalization of (\ref{eq:constitutive1}), whereas the
density-rate-conjugate microforce  balance provides the generalization of
(\ref{eq:mu}).

We articulate our basic laws of balance and imbalance in global form over a
generic subregion $\zcP$ of the region $\zcB$ occupied by the body in a
fixed reference configuration. The outward unit-normal field to
$\partial\zcP$, directed outward from $\zcP$, is designated by $\zn$.

\subsubsection{Atomic balance}
\label{subsect:mass}

We introduce the fields
\begin{center}
\begin{tabular}{cl}
{$\nu$}& \emph{atomic density}, \\ [0.5ex]
$\zjm$ & \emph{atomic flux}, \\[0.5ex]
$m$ & \emph{external atomic supply},
\\[0.5ex]
\end{tabular}
\end{center}
in which case the integrals
\begin{equation}
\int\limits_{\zcP}\nu\dv,
\qquad
\int\limits_{\partial\zcP}\zjm\!\cdot\!\zn\da,
\qquad\text{and}\qquad
\int\limits_{\zcP}m\dv
\end{equation}
represent, respectively, the number of atoms in $\zcP$, the number of atoms,
per unit time, added to $\zcP$ by diffusion across $\partial\zcP$, and the
number of atoms added, per unit time, to $\zcP$ by agencies external to
$\zcP$.\footnote{The supply $m$ should not be confused with a reaction
term. Rather, we view $m$ as an external supply that accounts for the insertion
of material by an external observer. Its role is analogous to that  of the
external heat supply that is routinely included in modern
continuum-thermodynamical statements of energy balance (see, for example,
Truesdell \cite{truesdell}). By introducing the external mass supply $m$, we
follow Gurtin \cite{G}.}

\emph{Atomic balance} is the postulate that, for each  region $\zcP$ and
each instant, the rate at which the number of atoms within $\zcP$ changes
with respect to time be equal to the rate at which atoms are added to $\zcP$
by the diffusion across $\partial\zcP$ and by external supplies to $\zcP$:
\begin{equation}
\zdot{\overline{\int\limits_{\zcP}\nu\dv}}
=-\int\limits_{\partial\zcP}\zjm\!\cdot\!\zn\da+\int\limits_{\zcP}m\dv.
\label{eq:mass_balance}
\end{equation}
Using the transport and divergence theorems, we obtain the equivalent local
field equation
\begin{equation}
\zzdot{\nu}=-\xd\zjm+m
\label{eq:local_mass_balance}
\end{equation}
enforcing atomic balance.

\subsubsection{Flux-conjugate microforce balance}
\label{subsect:micromoment}

Associated with the evolution of the atomic flux $\zjm$, we introduce a
system of power-conjugate \emph{microforces}, consisting of
\begin{center}
\begin{tabular}{cl}
$\zSigma$ & \emph{flux-conjugate microstress}, \\[0.5ex]
$\zh$     & \emph{flux-conjugate internal body microforce}, \\[0.5ex]
$\zell$   & \emph{flux-conjugate external body microforce},
\\[0.5ex]
\end{tabular}
\end{center}
in which case the integrals
\begin{equation}
\int\limits_{\zcP}\zSigma\zn\da,
\qquad
\int\limits_{\zcP}\zh\dv,
\qquad\text{and}\qquad
\int\limits_{\zcP}\zell\dv
\end{equation}
represent the flux-conjugate microforces exerted on the region $\zcP$ by the
flux-conjugate microtraction distributed over $\partial\zcP$, by agencies
within $\zcP$, and by flux-conjugate agencies external to $\zcP$,
respectively.

\emph{Flux-conjugate microforce balance} is the postulate that, for each
region $\zcP$ and each instant, the resultant, taking both internal and
external sources into account, of the flux-conjugate microforces acting on
$\zcP$ vanishes:
\begin{equation}
\int\limits_{\partial\zcP}\zSigma\zn\da
+\int\limits_{\zcP}(\zh+\zell)\dv={\bf0}.
\label{eq:orientational_balance}
\end{equation}
Using the divergence theorem, we arrive at the equivalent local field
equation
\begin{equation}
\xd\zSigma+\zh+\zell={\bf0}
\label{eq:local_orientational_balance}
\end{equation}
enforcing the flux-conjugate microforce balance.

\subsubsection{Density-rate-conjugate microforce balance}
\label{subsect:scalar}

Associated with the evolution of the atomic density $\nu$, we introduce a
system of power-conjugate \emph{microforces}, consisting of
\begin{center}
\begin{tabular}{cl}
{$\zxi$}& \emph{density-rate-conjugate microstress vector},     \\[0.5ex]
$\pi$ & \emph{density-rate-conjugate internal body microforce}, \\[0.5ex]
$\gamma$ & \emph{density-rate-conjugate external body microforce},
\\[0.5ex]
\end{tabular}
\end{center}
in which case the integrals
\begin{equation}
\int\limits_{\zcP}\zxi\!\zcdot\!\zn\da,
\qquad
\int\limits_{\zcP}\pi\dv,
\qquad\text{and}\qquad
\int\limits_{\zcP}\gamma\dv
\end{equation}
represent the density-rate-conjugate microforces exerted on the region $\zcP$
by the density-rate-conjugate microtraction distributed over $\partial\zcP$,
by agencies within $\zcP$, and by density-rate-conjugate agencies external to
$\zcP$, respectively.

\emph{Density-rate-conjugate microforce balance} is the postulate that, for
each region $\zcP$ and each instant, the resultant, taking both
internal and external sources into account, of the density-rate-conjugate
microforces acting on $\zcP$ vanishes:
\begin{equation}
\int\limits_{\partial\zcP}\zxi\!\zcdot\!\zn\da
+\int\limits_{\zcP}(\pi+\gamma)\dv=0.
\label{eq:scalar_balance}
\end{equation}
Using the divergence theorem, we arrive at the equivalent local field
equation
\begin{equation}
\xd\zxi+\pi+\gamma=0
\label{eq:local_scalar_balance}
\end{equation}
enforcing the density-rate-conjugate microforce balance.

\subsubsection{Energy imbalance}
\label{subsect:energy}

To formulate this imbalance, we introduce fields
\begin{center}
\begin{tabular}{cl}
{$\psi$}& \emph{free energy density}, \\[0.5ex]
$\mu$ &  \emph{diffusion potential}, \\[0.5ex]
\end{tabular}
\end{center}
in which case the integrals
\begin{equation}
\int\limits_{\zcP}\psi\dv
\qquad\text{and}\qquad
\int\limits_{\zcP}\mu\mskip1mu{m}\dv
\label{eq:orientational_diffusion}
\end{equation}
represent the free energy in $\zcP$ and rate at which energy is added to $\zcP$
through the external supply of atoms to $\zcP$.

Further, the integrals
\begin{equation}
\int\limits_{\partial\zcP}(\zSigma^{\trans}\zjm+\zxi\zzdot{\nu})\!\zcdot\!
\zn\da
\qquad\text{and}\qquad
\int\limits_{\zcP}(\zell\!\zcdot\!\zjm+\gamma\zzdot{\nu})\dv
\end{equation}
provide an accounting of the power expended on $\zcP$ by the tractions acting
on $\partial\zcP$ and by the agencies acting external to that region.

\emph{Energy imbalance} is the postulate that, for each region $\zcP$
and each instant:\footnote{To account for the diffusive flux of energy,
Gurtin \cite{G} included in his statement of energy imbalance the term
$$
-\int\limits_{\partial\zcP}\mu\zjm\!\zcdot\!\zn\da.
$$
Since, in our context, the diffusive flux of energy can be interpreted as the
surface power of the atomic diffusion, our theory accounts for this effect
through the term with integrand $\szSigma^{\trans}\zjm\!\zcdot\!\zn$.}
\begin{equation}
{\zdot{\overline{\int\limits_{\zcP}\psi\dv}}}
\le
\int\limits_{\partial\zcP}(\zSigma^{\trans}\zjm+\zxi\zzdot{\nu})\!\zcdot\!
\zn\da
+\int\limits_{\zcP}(\zell\!\zcdot\!\zjm+\gamma\zzdot{\nu}+\mu\mskip1mu{m})\dv.
\label{eq:difference}
\end{equation}
Using the transport and divergence theorem as before, we arrive at the local
inequality
\begin{equation}
-\skew4\dot{\psi}
+(\mu-\pi)\zzdot{\nu} +\zxi\!\zcdot \! \xg\zzdot{\nu}
-\zh\!\zcdot\!\zjm
+(\zSigma+\mu\idem)
\!\zcdot\!\xg\zjm
\ge0
\label{eq:local_strong_energy_imbalance_2}
\end{equation}
enforcing energy imbalance.

\subsection{Constitutive equations}
\label{sect:const}

We assume that the free energy density $\psi$, the flux-congugate microstress
$\zSigma$, the flux-congugate internal body microforce $\zh$, the
density-rate-conjugate microstress $\zxi$, and the density-rate-conjugate
internal scalar body microforce $\pi$ are determined constitutively in terms
of the atomic density $\nu$, the gradient $\xg\nu$ of the atomic density, the
diffusion potential $\mu$, the gradient $\xg\mu$ of the diffusion potential,
and the atomic flux $\zjm$:
\begin{equation}
\left.
\begin{split}
\psi&=\hat{\psi}(\nu,\xg\nu,\mu,\xg\mu,\zjm),
\\
\zSigma&=\phat{\zSigma}(\nu,\xg\nu,\mu,\xg\mu,\zjm),
\\
\zh&=\phat{\zh}(\nu,\xg\nu,\mu,\xg\mu,\zjm),
\\
\zxi&=\phat{\zxi}(\nu,\xg\nu,\mu,\xg\mu,\zjm),
\\
\pi&=\hat{\pi}(\nu,\xg\nu,\mu,\xg\mu,\zjm).
\end{split}
\right\}
\label{eq:cr}
\end{equation}
We emphasize that, in contrast to the standard approach to the theory of
atomic diffusion, neither the diffusion potential $\mu$ nor the atomic flux
$\zjm$ is given by a constitutive relation.

Inserting the constitutive relations (\ref{eq:cr}) into the local energy
imbalance (\ref{eq:local_strong_energy_imbalance_2}) and writing, for
brevity,
\begin{equation}
z=(\nu,\xg\nu,\mu,\xg\mu,\zjm),
\end{equation}
we arrive at the functional inequality
\begin{multline}
\bigl(\frac{\partial\hat{\psi}}{\partial\nu}(z)
-{\mu}+\phat{\pi}(z)\bigr)\zzdot{\nu}
+\bigl(\frac{\partial\hat{\psi}}{\partial(\xg\nu)}(z)
-\phat{\zxi}(z)\bigr)\!\zcdot\!\xg\zzdot{\nu}
+\frac{\partial\hat{\psi}}{\partial\mu}(z)\zzdot{\mu}
\\
+\frac{\partial\hat{\psi}}{\partial(\xg\mu)}(z)\!\zcdot\!\xg\zzdot{\mu}
+\frac{\partial\hat{\psi}}{\partial\zjm}(z)
\!\zcdot\!\zzdot{\zjm}
+\phat{\zh}(z)\!\zcdot\!\zjm 
-\bigl(\phat{\zSigma}(z)+{\mu}\idem\bigr)
\!\zcdot\!\xg\zjm
\le0.
\label{eq:dissipation_imbalance_with_constitutive_relations}
\end{multline}
Hence, following the procedure founded by Coleman \& Noll \cite{CN} in their
incorporation of the second law into continuum thermomechanics, we find that:
\begin{itemize}
\item[({\em i\/})] the constitutive response functions $\hat{\psi}$ and
$\phat{\zxi}$ delivering the free energy density $\psi$ and the
density-rate-conjugate microstress $\zxi$ must be independent of the
diffusion potential $\mu$, the gradient $\xg\mu$ of the diffusion potential,
and the atomic flux $\zjm$, and obey
\begin{equation}
\hat{\zxi}(\nu,\xg\nu)
=\frac{\partial\phat{\psi}}{\partial(\xg\nu)}(\nu,\xg\nu);
\label{eq:basic_model_restrictions_1}
\end{equation}
\item[({\em ii\/})] the constitutive response function $\phat{\zSigma}$
delivering the flux-conjugate microstress $\zSigma$ must be independent of
the atomic density $\nu$, the gradient $\xg\nu$ of the atomic density, the
gradient $\xg\mu$ of the diffusion potential, and the atomic flux $\zjm$,
and obey
\begin{equation}
\phat{\zSigma}=-\mu\idem;
\label{eq:basic_model_restrictions_2}
\end{equation}
\item[({\em iii\/})] the constitutive response function $\hat{\pi}$
delivering the density-rate-conjugate internal body microforce $\pi$ must be
independent of the gradient $\xg\mu$ of the diffusion potential and the
atomic flux $\zjm$, and obey
\begin{equation}
\hat{\pi}(\nu,\xg\nu,\mu)=
\mu-\frac{\partial\hat{\psi}}{\partial\nu}(\nu,\xg\nu);
\label{eq:basic_model_restrictions_3}
\end{equation}
\item[({\em iv\/})] the constitutive response function $\hat{\zh}$ for the
flux-conjugate internal body microforce $\zh$ must be consistent with the
\emph{residual inequality}
\begin{equation}
\phat{\zh}(\nu,\xg\nu,\mu,\xg\mu,\zjm)\!\zcdot\!\zjm\le0.
\label{eq:residual_inequality}
\end{equation}
\end{itemize}

Further, granted smoothness of the response function $\phat{\zh}$, a result
due to Gurtin \& Voorhees \cite{gurtin93} yields a general solution
\begin{equation}
\zh=-\zZ(\nu,\xg\nu,\mu,\xg\mu,\zjm)\zjm
\label{eq:solution_of_residual_inequality}
\end{equation}
of (\ref{eq:residual_inequality}), where the \emph{reciprocal mobility}
tensor $\zZ$  must obey
\begin{equation}
\zjm\!\zcdot\!\zZ(\nu,\xg\nu,\mu,\xg\mu,\zjm)\zjm\ge0.
\end{equation}

The governing equations that arise on substituting the foregoing
thermodynamically consistent constitutive equations in the local field
equations (\ref{eq:local_scalar_balance}) and
(\ref{eq:local_orientational_balance}) expressing flux-conjugate microforce
balance and density-rate-conjugate microforce balance read
\begin{equation}
\left\delimiter0
\begin{split}
%
\zZ(\nu,\xg\nu,\mu,\xg\mu,\zjm)\,\zjm+\xg\mu&=\zell,
\\[4pt]
\frac{\partial\phat{\psi}}{\partial\nu}(\nu,\xg\nu)
-\xd\bigl(\frac{\partial\phat{\psi}}{\partial(\xg\nu)}(\nu,\xg\nu)\bigr)-\mu&
=\gamma.
\quad
\end{split}
\right\}
\label{eq:basic_model_final_equations}
\end{equation}
Together with the atomic balance (\ref{eq:local_mass_balance}),
(\ref{eq:basic_model_final_equations}) form the final governing equations of
our theory for the diffusion of a single unconstrained atomic species on a
fixed lattice.

The foregoing results show that the behavior of a medium of the sort
considered here is completely determined by the provision of two
constitutive response functions:
\begin{itemize}
\item[$\bullet$] $\hat{\psi}$ determining the free energy density as a
function of
the density $\nu$ and its gradient $\xg\nu$; and
\item[$\bullet$] the  reciprocal mobility tensor $\zZ$, which,
in general, may depend on $\nu$, $\xg\nu$, $\mu$, $\xg\mu$, and $\zjm$.
\end{itemize}

Provided that the external supplies $m$, $\zell$, and $\gamma$ all vanish,
that the constitutive function  $\phat{\psi}$ is quadratic in the gradient
of the atomic density and twice differentiable with respect to the atomic
density, and that $\zZ=\zeta\idem$ with $\zeta$ depending at most on the
atomic density, the system formed by combining
(\ref{eq:basic_model_final_equations}) with the atomic balance
(\ref{eq:local_mass_balance}) yields precisely the Cahn-Hilliard equation
(\ref{eq:cahn-hilliard}). If, furthermore, $\hat{\psi}$ is independent of the
gradient of the atomic density, we obtain
\begin{equation}
\zjm=
-\frac{\hat{\psi}{}^{\prime\prime}(\nu)}{\zeta(\nu)}\xg\nu,
\label{eq:expression_for_flux}
\end{equation}
for the atomic flux $\zjm$ in terms of the density $\nu$ and its gradient
$\xg\nu$. Comparing (\ref{eq:expression_for_flux}) with the conventional
relation (\ref{eq:constitutive2}), we obtain an expression
\begin{equation}
D(\nu)=\frac{\hat{\psi}{}^{\prime\prime}(\nu)}{\zeta(\nu)}
\label{diffusivity}
\end{equation}
for the diffusivity. Hence, within our framework, a constitutive relation of
the form (\ref{eq:constitutive2}) can, therefore, be interpreted as an
expression of flux-conjugate microforce balance, granted that the
flux-conjugate internal body microforce is isotropic and linear in the atomic
flux and that the flux-conjugate external body microforce vanishes. If,
moreover, we assume that $\hat{\psi}$ has the particular form
\begin{equation}
\hat{\psi}(\nu)=\kBT\mskip1mu\nu\mskip1mu\log\nu
\label{eq:ideal}
\end{equation}
corresponding to the classical entropic contribution to the free energy of a
dilute mixture, with $\kB$ \emph{Boltzmann's constant} and $\theta$ the
\emph{absolute temperature}, then (\ref{diffusivity}) reduces to
\begin{equation}D(\nu)=\frac{\kBT}{\zeta(\nu)},
\label{eq:stokes_einstein}
\end{equation}
the standard relation between diffusivity and mobility (see, for example,
Nernst \cite{nernst88} and Einstein \cite{einstein05}). More generally,
granted that the reciprocal mobility tensor $\zZ$ is invertible, our theory
in conjunction with the assumption (\ref{eq:ideal}) yields the diffusivity
as a tensor
\begin{equation}
\zD(\nu,\xg\nu,\mu,\xg\mu,\zjm)=\kBT{\zZ}^{-1}(\nu,\xg\nu,\mu,\xg\mu,\zjm).
\end{equation}

\section{Theory accounting for both atomic diffusion and lattice deformation}
\label{sect:generalization}

\subsection{Balance and imbalance laws}

In addition to the mass balance (\ref{eq:mass_balance}), flux-conjugate
microforce balance (\ref{eq:orientational_balance}), and
density-rate-conjugate microforce balance (\ref{eq:scalar_balance}), the
theory is based upon the following laws:
\begin{itemize}
\item[$\bullet$] lattice-velocity-conjugate force balance;
\item[$\bullet$] lattice-velocity-conjugate torque balance;
\item[$\bullet$] energy imbalance.
\end{itemize}
Again, the standard formulation of diffusion in a deformable body is
supplemented with two microforce balances, which will provide appropriate
generalizations of (\ref{eq:constitutive1}) and  (\ref{eq:mu}).

\subsubsection{Lattice-velocity-conjugate force balance and
lattice-velocity-conjugate torque balance}

The motion of the lattice is described by a mapping $\zy$ of points $\zx$
in $\zcB$ and times $t$ into points $\zy(\zx,t)$ of space. We require that
$\Det(\xg\zy)>0$. Associated with this motion, we introduce a system of
power-conjugate forces consisting of
\begin{center}
\begin{tabular}{cl}
$\zT$&  \emph{lattice-velocity-conjugate} (\emph{Piola}) \emph{stress},
\\[0.5ex]
$\zb$ & \emph{lattice-velocity-conjugate body force},
\\[0.5ex]
\end{tabular}
\end{center}
in which case the integrals
\begin{equation}
\int\limits_{\zcP}\zT\zn\da
\qquad\text{and}\qquad
\int\limits_{\zcP}\zb\dv
\end{equation}
represent the lattice-velocity-conjugate forces exerted on the region $\zcP$
by the lattice-velocity-conjugate traction distributed over $\partial\zcP$
and by lattice-velocity-conjugate agencies external to $\zcP$, respectively.
Further, the integrals
\begin{equation}
\int\limits_{\zcP}\zy\!\times\!\zT\zn\da
\qquad\text{and}\qquad
\int\limits_{\zcP}\zy\!\times\!\zb\dv
\end{equation}
represent the lattice-velocity-conjugate torques exerted on the region $\zcP$
by the lattice-velocity-conjugate traction distributed over $\partial\zcP$
and by lattice-velocity-conjugate agencies external to $\zcP$, respectively.

\emph{Lattice-velocity-conjugate force balance} is the postulate that, for
each region $\zcP$ and each instant, the resultant of the
lattice-velocity-conjugate forces acting on $\zcP$ vanishes:
\begin{equation}
\int\limits_{\partial\zcP}\zT\zn\da
+\int\limits_{\zcP}\zb\dv={\bf0}.
\label{eq:lattice_balance_1}
\end{equation}
Further, \emph{lattice-velocity-conjugate torque balance} is the postulate
that, for each region $\zcP$ and each instant, the resultant of the torques
associated with the lattice-velocity-conjugate forces acting on $\zcP$
vanishes:
\begin{equation}
\int\limits_{\partial\zcP}\zy\!\times\!\zT\zn\da
+\int\limits_{\zcP}\zy\!\times\!\zb\dv={\bf0}.
\label{eq:lattice_balance_2}
\end{equation}
In writing (\ref{eq:lattice_balance_2}), we have assumed that the diffusing
atoms do not induce torques, as would be expected at least for approximately
spherical atoms. Using the divergence theorem, we arrive at the equivalent
local field equations
\begin{equation}
\xd\zS+\zb={\bf0}
\qquad\text{and}\qquad
(\xg\zy)\zT^{\trans}=\zT(\xg\zy)^{\trans}.
\label{eq:local_cauchy_balance}
\end{equation}

\subsubsection{Energy imbalance}

To account for power expenditures associated with the motion of the lattice,
we modify the energy imbalance (\ref{eq:difference}) to read
\begin{equation}
{\zdot{\overline{ \int\limits_{\zcP}
\psi\dv }}}
\ \le\
\int\limits_{\partial\zcP}\bigl(
\zSigma^{\trans}\zjm+\zxi\zzdot{\nu}+\zT^{\trans}\skew3\dot{\zy}
\bigr)\!\zcdot\!\zn\!\da
+\int\limits_{\zcP}(
\zl\!\zcdot\!\zjm+\pi\zzdot{\nu}+\mu\mskip1mu{m}+\zb\!\zcdot\!
  \skew3\dot{\zy})\dv.
\label{eq:difference2}
\end{equation}
Proceeding as before, we arrive at the local inequality
\begin{multline}
-\dot{\psi}
+(\mu-\pi)\zzdot{\nu}+\zxi\!\zcdot\!\xg\zzdot{\nu}
-\zh\!\zcdot\!\zjm
\\
+\zT\!\zcdot\!\xg\skew3\dot{\zy}+ (\zSigma+\mu\idem)
\!\zcdot\!\xg\zjm \ \ge \ 0.
\label{eq:delta2}
\end{multline}

\subsection{Constitutive equations}

We add the lattice-velocity-conjugate stress and the lattice-motion gradient
$\xg\zy$ to the lists of dependent and independent constitutive variables,
respectively, so that
\begin{equation}
\left.
\begin{split}
\psi&=\phat{\psi}(\nu,\xg\nu,\mu,\xg\mu,\zjm,\xg\zy),
\\
\zSigma&=\phat{\zSigma}(\nu,\xg\nu,\mu,\xg\mu,\zjm,\xg\zy),
\\
\zh&=\phat{\zh}(\nu,\xg\nu,\mu,\xg\mu,\zjm,\xg\zy),
\\
\zxi&=\phat{\zxi}(\nu,\xg\nu,\mu,\xg\mu,\zjm,\xg\zy),
\\
\pi&=\phat{\pi}(\nu,\xg\nu,\mu,\xg\mu,\zjm,\xg\zy),
\\
\zT&=\phat{\zT}(\nu,\xg\nu,\mu,\xg\mu,\zjm,\xg\zy).
\end{split}
\right\}
\label{eq:cr2}
\end{equation}

Inserting the constitutive relations (\ref{eq:cr2}) into the local energy
imbalance (\ref{eq:delta2}) and writing
\begin{equation}
z=(\nu,\xg\nu,\mu,\xg\mu,\zjm,\xg\zy),
\end{equation}
we arrive at the functional inequality
\begin{multline}
\bigl(\frac{\partial}{\partial\nu}\hat{\psi}(z)
-\mu+\phat{\pi}(z)\bigr)\zzdot{\nu}
+\bigl(\frac{\partial\hat{\psi}}{\partial(\xg\nu)}(z)-\hat{\zxi}(z)\bigr)
   \!\zcdot\!\xg\zzdot{\nu}
+\frac{\partial\hat{\psi}}{\partial\mu}(z)\zzdot{\mu}
\\
+\frac{\partial\hat{\psi}}{\partial(\xg\mu)}(z)\!\zcdot\!\xg\zzdot{\mu}
+\bigl(\frac{\partial\hat{\psi}}{\partial(\xg\zy)})(z)
-\phat{\zT}(z)\bigr)\!\zcdot\!\xg\zzdot{\zy}
\\
+\frac{\partial\hat{\psi}}{\partial\zjm}(z)\!\zcdot\!\zzdot{\zjm}
%
+\phat{\zh}(z)\!\zcdot\!\zjm
-\bigl(\phat{\zSigma}(z)+\mu\idem\bigr)
\!\zcdot\!\xg\zjm\ \le\ 0.
\label{eq:dissipation_imbalance_with_constitutive_relations2}
\end{multline}
Hence, proceeding as before, we obtain the results
\begin{itemize}
\item[({\em i\/})] the constitutive response functions $\hat{\psi}$,
$\phat{\zT}$, and $\phat{\zxi}$ delivering the energy density $\psi$, the
lattice-velocity-conjugate stress ${\zT}$,  and the density-rate-conjugate
microstress $\zxi$ must be independent of the diffusion potential $\mu$, the
gradient $\xg\mu$ of the diffusion potential, and the atomic flux $\zjm$, and
obey
\begin{equation}
\left.
\begin{split}
\phat{\zT}(\nu,\xg\nu,\xg\zy)
&=\frac{\partial\phat{\psi}}{\partial(\xg\zy)}(\nu,\xg\nu,\xg\zy),
\\[4pt]
\phat{\zxi}(\nu,\xg\nu,\xg\zy)
&=\frac{\partial\phat{\psi}}{\partial(\xg\nu)}(\nu,\xg\nu,\xg\zy);
\end{split}
\right\}
\label{eq:basic_model_restrictions_deformation_1}
\end{equation}
\item[({\em ii\/})] the constitutive response function $\phat{\zSigma}$
delivering the flux-conjugate microstress $\zSigma$ must be independent of
the atomic density $\nu$, the gradient $\xg\nu$ of the atomic density, the
gradient $\xg\mu$ of the diffusion potential, and the atomic flux $\zjm$,
and obey
\begin{equation}
\phat{\zSigma}=-\mu\idem;
\label{eq:basic_model_restrictions_deformation_2}
\end{equation}
\item[({\em iii\/})] the constitutive response function $\hat{\pi}$
delivering the density-rate-conjugate internal body microforce $\pi$ must be
independent of the gradient $\xg\mu$ of the diffusion potential and the
atomic flux $\zjm$, and obey
\begin{equation}
\hat{\pi}(\nu,\xg\nu,\mu)=
\mu-\frac{\partial\hat{\psi}}{\partial\nu}(\nu,\xg\nu);
\label{eq:basic_model_restrictions_deformation_3}
\end{equation}
\item[({\em iv\/})] the constitutive response function $\hat{\zh}$ for the
flux-conjugate internal body microforce $\zh$ must be consistent with the
\emph{residual inequality}
\begin{equation}
\phat{\zh}(\nu,\xg\nu,\mu,\xg\mu,\zjm,\xg\zy)\!\zcdot\!\zjm\le0.
\label{eq:residual_inequality_deformation}
\end{equation}
\end{itemize}

Analogous to the result (\ref{eq:solution_of_residual_inequality}),
we have the general solution
\begin{align}
\zh=&-\zZ(\nu,\xg\nu,\mu,\xg\mu,\zjm,\xg\zy)\,\zjm
\label{eq:solution_of_residual_inequality_deformation}
\end{align}
of (\ref{eq:residual_inequality_deformation}), where the
\emph{reciprocal mobility} tensor $\zZ$ must obey
\begin{equation}
\zjm\!\zcdot\!\zZ(\nu,\xg\nu,\mu,\xg\mu,\zjm)\zjm\ge0.
\end{equation}

Further, the constitutive relations are restricted by objectivity; that
is, under the observer transformation
\begin{equation}
\left.
\begin{split}
\nu&\mapsto\nu,
\\
\mu&\mapsto\mu,
\end{split}
\qquad
\begin{split}
\xg\nu&\mapsto\xg\nu,
\\
\xg\mu&\mapsto\xg\mu,
\end{split}
\qquad
\begin{split}
\zjm&\mapsto\zjm,
\\
\xg\zy&\mapsto\zQ(\xg\zy),
\end{split}
\right\}
\end{equation}
with $\zQ$ orthogonal, we must have
\begin{equation}
\left.
\begin{split}
\zSigma&\mapsto\zQ\zSigma,
\\
\zxi&\mapsto\zxi,
\\
\zT&\mapsto\zQ\zT,
\end{split}
\qquad
\begin{split}
\zh&\mapsto\zh,
\\
\pi&\mapsto\pi,
\\
\zb&\mapsto\zb,
\end{split}
\qquad
\begin{split}
\zell&\mapsto\zell,
\\
\gamma&\mapsto\gamma,
\\
\psi&\mapsto\psi.
\end{split}
\right\}
\end{equation}
Granted the foregoing thermodynamic results, these transformations require,
in particular, that
\begin{equation}
\left.
\begin{split}
\phantom{\frac{\partial}{\partial}}
\psi&=\tilde{\psi}(\nu,\xg\nu,\zC),
\\
\phantom{\frac{\partial}{\partial}}
\zZ&=\skew3\tilde{\zZ}(\nu,\xg\nu,\mu,\xg\mu,\zjm,\zC),
\\
\zT&=(\xg\zy)\frac{\partial\tilde{\psi}}{\partial\zC}(\nu,\xg\nu,\zC),
\end{split}
\right\}
\label{eq:cr3}
\end{equation}
with $\zC=(\xg\zy)^{\trans} \,\xg\zy$ the \emph{right Cauchy-Green tensor}.

The governing equations that arise on substituting the foregoing
thermodynamically consistent constitutive equations in the local field
equations (\ref{eq:local_cauchy_balance})$_1$,
(\ref{eq:local_scalar_balance}) and
(\ref{eq:local_orientational_balance}) expressing lattice-velocity-conjugate
force balance, flux-conjugate microforce balance, and
density-rate-conju\-gate microforce balance read
\begin{equation}
\left\delimiter0
\begin{split}
%
\xd\bigl((\xg\zy)
\frac{\partial\tilde{\psi}}{\partial\zC}(\nu,\xg\nu,\zC)\bigl)+\zb&={\bf0},
\\[4pt]
\skew3\tilde{\zZ}(\nu,\xg\nu,\mu,\xg\mu,\zjm,\zC)\,\zjm+\xg\mu&=\zell,
\\[4pt]
\frac{\partial\tilde{\psi}}{\partial\nu}(\nu,\xg\nu,\zC)
-\xd\bigl(\frac{\partial\tilde{\psi}}{\partial(\xg\nu)}(\nu,\xg\nu,\zC)\bigr)
-\mu&=\gamma.
\end{split}
\right\}
\label{eq:deformation_model_final_equations}
\end{equation}
Together with the atomic balance (\ref{eq:local_mass_balance}),
(\ref{eq:deformation_model_final_equations}) form the final governing
equations of our theory for the diffusion of a single unconstrained atomic
species on a deformable lattice lattice.

We find that the behavior of a medium of the sort considered here is still
completely determined by the provision of two constitutive response
functions:
\begin{itemize}
\item[$\bullet$] $\tilde{\psi}$ determining the energy density as a function
of the density $\nu$, the gradient $\xg\nu$ of the atomic density, and right
Cauchy-Green tensor $\zC$ associated with the lattice deformation; and
\item[$\bullet$] the reciprocal mobility tensor $\skew3\tilde{\zZ}$,
which, in general, may depend on the atomic density $\nu$, the gradient
$\xg\nu$ of the atomic density, the diffusion potential $\mu$, the gradient
$\xg\mu$ of the diffusion potential, the atomic flux $\zjm$, and the right
Cauchy-Green tensor $\zC$ associated with the lattice deformation.
\end{itemize}

\section{Discussion }
\label{sect:discussion}

Our theory is based upon introducing generalized velocities for diffusion
and corresponding power-conjugate forces, which we call microforces, into an
otherwise conventional continuum-mechanical description. In particular, we
identify two distinct generalized velocities---a vector-valued diffusive
atomic flux that characterizes simple diffusion and  a scalar-valued atomic
density rate that characterizes phase separation. A systematic,
thermodynamically consistent derivation  generates a final system of
governing  equations---(\ref{eq:local_mass_balance}) and
(\ref{eq:basic_model_final_equations}) for a fixed lattice, and
(\ref{eq:local_mass_balance}) and
(\ref{eq:deformation_model_final_equations}) for a deformable lattice---that
constitutes a generalization of common diffusion equations, a generalization
that accounts for constitutive nonlinearities and external forces acting on
the diffusing atoms.

We find that the constitutive response of the material is determined by the
provision of two  functions, one for the energy density and the other for
the reciprocal mobility tensor. In particular, the Cahn-Hilliard equation
(\ref{eq:cahn-hilliard}) results on assuming that the energy response
function is quadratic in $\xg\nu$ and that the reciprocal mobility tensor
is proportional to the identity tensor by a scalar coefficient that is at
most a function of $\nu$. Importantly, we arrive at our theory without
introducing a constitutive relation such as (\ref{eq:constitutive1}) for the
diffusive flux.

Additionally, our proposed format also leads to a generalization,
(\ref{diffusivity}), of the classical relation between diffusion coefficient
and viscosity. We find that the classical form for the  relation
(\ref{eq:stokes_einstein}) holds  when the  free energy is given by the
special constitutive relation (\ref{eq:ideal}).

A distinction between the final governing equations of our theory and those
arising in more conventional derivations is that, in our approach, a
vector-valued microforce balance provides a generalization between diffusive
atomic flux and gradient of the diffusion potential.

\section*{Acknowledgements}

This work was performed with financial support from the U.\ S.\ Department of
Energy and the EPSRC (U.K.).


\end{document}